# EFFICIENT LIVER SEGMENTATION WITH 3D CNN USING COMPUTED TOMOGRAPHY SCANS

Khaled Humady[1], Yasmeen Al-Saeed[2,3], Nabila Eladawi[4], Ahmed Elgarayhi[1], Mohammed Elmogy[2, *], Mohammed Sallah[1,5]

[1]Applied Mathematical Physics Research Group, Physics Department, Faculty of Science, Mansoura University, Mansoura 35516, Egypt
[2]Information Technology Department, Faculty of Computers and Information, Mansoura University, Mansoura 35516, Egypt
[3]Information Technology Department, Faculty of Computers and Artificial Intelligence, South Valley University, Hurghada 84511, Egypt.
[4]Information Systems Department, Faculty of Computers and Information, Mansoura University, Mansoura 35516, Egypt
[5]Higher Institute of Engineering and Technology, New Damietta City, Egypt

Corresponding author: Mohammed Elmogy (melmogy@mans.edu.eg)

The first two authors contributed equally to this work and shared the first authorship.

The fifth and sixth authors are sharing the senior authorship.

## ABSTRACT

The liver is one of the most critical metabolic organs in vertebrates due to its vital functions in the human body, such as detoxification of the blood from waste products and medications. Liver diseases due to liver tumors are one of the most common mortality reasons around the globe. Hence, detecting liver tumors in the early stages of tumor development is highly required as a critical part of medical treatment. Many imaging modalities can be used as aiding tools to detect liver tumors. Computed tomography (CT) is the most used imaging modality for soft tissue organs such as the liver. This is because it is an invasive modality that can be captured relatively quickly. This paper proposed an efficient automatic liver segmentation framework to detect and segment the liver out of CT abdomen scans using the 3D CNN DeepMedic network model. Segmenting the liver region accurately and then using the segmented liver region as input to tumors segmentation method is adopted by many studies as it reduces the false rates resulted from segmenting abdomen organs as tumors. The proposed 3D CNN DeepMedic model has two pathways of input rather than one pathway, as in the original 3D CNN model. In this paper, the network was supplied with multiple abdomen CT versions, which helped improve the segmentation quality. The proposed model achieved 94.36%, 94.57%, 91.86%, and 93.14% for accuracy, sensitivity, specificity, and Dice similarity score, respectively. The experimental results indicate the applicability of the proposed method.

## 1. INTRODUCTION

According to the world health organization (WHO), cancer is the leading cause of death worldwide and is typically accountable for at least one death in every six deaths. About 30% of cancer cases in developing countries are caused by hepatitis or the papilloma virus (HPV). Many cancers can be cured if detected and treated properly in the early stages. Generally, individuals with chronic liver illness had a worse health-related quality of life (HRQL) than the general population where similar HRQL to those with congestive heart failure and chronic or obstructive lung disease [1].

The liver is one of the most critical metabolic organs in the human body [2]. No biochemical test can well characterize the liver's functioning since the organ conducts many synthetic, biochemical, and excretory processes. Despite receiving a lot of criticism for this, the term "liver function tests" is deeply embedded in medical jargon. It may be argued that "liver damage testing" would be a better term.

Additionally, it is important to consider the clinical history while evaluating the functions. The significance of several disease markers in radiological imaging cannot be overstated [3]. Primary liver diseases with or without icterus include infectious hepatitis, toxic necrosis, hepatic hemangiomas, supportive hepatitis, hepatica, bile duct adenoma, malignancy, chronic active hepatitis, and copper storage disease. Some secondary liver illnesses are infiltrative lip doses linked to hypothyroidism, diabetes, pancreatitis, malnutrition, prolonged passive congestion in cardiovascular decompensation, distant metastatic hepatic cancers, and supplemental amyloidosis. The three conditions of hemolytic crises, intrahepatic biliary liver failure, and extrahepatic bile duct obstruction, should not be confused with icterus [4].

There are many liver disorders, for instance, chronic hepatitis B virus (HBV) infection, which is a significant global cause of cirrhosis and hepatocellular cancer [5]. Chronic hepatitis C (CH-C), a common and potentially fatal liver condition, is now being treated with new therapies that have better cure rates, less side effects, and shorter treatment times [6]. Alcoholic hepatitis can be differentiated from early ASH (advanced phases of nonalcoholic fatty liver disease) in fully compensated patients since it can be the first symptom of clinically silent Adrenoleukodystrophy, a fatal genetic disease, or as an exacerbation of preexisting cirrhosis: severe sepsis, biliary blockage, diffuse HCC, and drug-induced liver disease [7].

The most frequent cause of mortality in people is a cardiovascular disease rather than hepatic illness, while NASH patients have a higher chance of dying from liver-related causes [8]. In addition, multiple fluid-filled liver cysts are a defining feature of polycystic liver disease. PLD can be a secondary symptom of autosomal dominant polycystic liver failure, a predominant symptom of autosomal dominant polycystic kidney disease, or appear in the context of two different inherited illnesses. Each has different follow-up, counseling, family screening, and prognosis [9].

One of the most prevalent dangerous medical conditions that cause mortality globally is liver cancer [10]. . Early detection and accurate evaluation of cancers are therefore essential. The most successful therapy is surgical excision; however, it could only considered with some types of tumors. With most tumor types, CT imaging is considered a great resolution that resulted in high detection rates. However, differentiating liver tumors might be challenging due to the appearance similarity between healthy and tumor tissues in most cases [11]. On non-enhanced CT scans, HCC typically appears as a hypo attenuating mass. It could exhibit lesser attenuation in the center, indicating necrosis or bleeding inside the tumor. Additionally, fat attenuation might exist. Hepatocellular carcinoma is often highly vascular and in most instances enhanced on arterial phase imaging [12].

Benign lesions result from any abnormality in the structure of stromal and epithelial cells that make up the liver. The most common benign liver lesions are; hepatic hemangioma which develops from endothelium cells, along with cyst lesions produced from biliary epithelium. Biliary epithelium is typically caused by discernible reason, such as the use of oral contraceptives, anabolic steroids, or a disorder affecting glycogen storage. In general, throughout the years of childbearing, women are afflicted more frequently than males [13]. The lesion's cellular characteristics and trabecular structure are often closer to normal grossly than histologically, making it easier to discern the lesion's boundaries. Typically, there is no fibrous capsule. An abrupt transformation of the portal structures might point to the margin. Fat and glycogen may be present in tumor cells [14].

The usage of artificial intelligence (AI) techniques can provide accurate, efficient results without much concern about human biases, making it one of the most significant technological advances in the foreseeable future [15]. In recent years, the early detection, diagnosis, and management of diseases have benefited severely from the usage of several medical imaging methods, such as computed tomography (CT). In clinical settings, the interpretation and analysis of medical images are essentially the responsibility of radiologist [16]. The usage of AI techniques can complement the radiologist abilities to acquire a significant amount of information from medical imaging, which might improve the precision of diagnoses. Due to factors related to AI method implementation for CT scans analysis such as as data form, CT slices density, training duration, and data augmentation level, AI methods efficient is questioned [17]. AI is usually used with visual data from of CT scans to segment, categorize lesions such as liver tumors, and to estimate prognosis. Hepatic AI-assisted diagnosis aim to detect, characterize and estimate prognosis in patients with hepatic cancer [18].

Segmenting the liver area accurately and then using the segmented liver region rather than the whole abdomen CT scan as input to the tumors segmentation algorithm can be beneficial in reducing error rates resulting from segmenting tumors-like abdomen organs. Hence, improving the quality of medical diagnosis. The main contribution of this paper is as follows. First, enhancing the perceptibility of abdomen CT scans in terms of the liver by focusing on the liver region intensities rather than the whole abdomen region using Hounsfield units (HU) scaling, suppressing impulse noises from the CT scanner along with enhancing the contrast of the CT scan. The second contribution of this paper is providing an efficient automatic liver segmentation framework based on a 3D CNN DeepMedic network to segment the liver out of other abdomen organs as a pre-stage for tumor segmentation. The output of the proposed method can be used as input to the tumor segmentation method.

The rest of this paper is organized as follows. Section 2 discusses the related work. Section 3 presents the methodology by which the proposed framework was constructed. Section 4 presents the experimental results along with a discussion. Finally, the conclusion and future work is presented in Section 5.

## 2. RELATED WORK

Incorporating AI in medical image analysis attract many research studies. To evaluate the potential use of a computer-aided diagnosis system (CAD) with texture classification in the differential diagnosis of liver malignancies, Huang et al. [19] presented non enhanced CT, where the tumor spot was manually located and extracted from the scanned CT image as a circle. The sub image's auto covariance texture features were extracted, and SVM technique was used to determine if the tumor was benign or malignant.

A unique method and technique for segmenting the liver and its inner lesions using CT images were proposed by Laurent and Sergio [20]. There is no interaction between the user and the analytical system during startup because the procedure is fully automated. They segmented the liver tissue out of other abdominal organs, using statistical models. They used with an active contour approach to utilize and refine liver surface segmentation. After that, liver parenchyma and hepatic lesions were labeled using classification.

A hybrid method that can automatically divide the liver from an abdominal CT scan and find hepatic anomalies was presented by Ahmed et al. [21]. They could compute the region of the liver affected as a tumor lesion and count the discernible lesions. This method was one of the quickest and yet most accurate technique to test the presence of liver tumors. They concluded that their method for segmenting liver lesions from the patient database could achive high reliability level.

A CBIR-based CAD system was proposd by Peter et al. [22], which reliably distinguished liver lesions and has shown potential accuracy even for challenging-to-interpret, sub-centimeter hepatic lesions. The proposed method prioritizes radiological usefulness while being easily incorporated into regular activities. The software program outperformed doctors when diagnosing liver cancers.

In order to identify carcinoma and hemangioma in human liver tumors, Kumar et al. [23] established texture based algorithms to examine tumor textures. Different texture-modeling methods were used: two traditional grey level techniques (GLT), a contemporary wavelet (WCT) and contour lets method (CLT). Three strategies were compared regarding detection accuracy levels and other performance indicators. The CLT features outperformed the GLT and WCT features in general for the data set under consideration. The performance to evaluate whether a tumor is malignant or benign was evaluated to have the beat results with CLT characteristics.

An automated CAD system that can differentiate tumors, cysts, stones, and healthy tissues was developed by Gaurav and Saini [24]. The wavelet and curve let multiresolution techniques were used in their approach. The most valuable features were selected using a genetic algorithm from the list of retrieved qualities. Then, SVM and artificial neural networks (ANN) were used as classifiers for classification.

Hariharan [25] suggested using AI to enhance the perceptibility of obtained medical image and segment organs based on thresholding technique. To get the ideal threshold value for accurate segmentation, the input scan was adaptively thresholding using a fuzzy-based Shannon's entropy function. Patrick et al. [26] offered the license that allowed the reserechers to use CT data for additional medical purposes.

Das et al. [27] suggested a hybrid approach which incorporated fuzzy clustering and adaptive thresholding for segmenting CT abdominal images. According to the imaging intensity of the tissues, the tumor region was identified using fuzzy-clustering approach. An adaptive thresholding was used to distinguish the liver from the surrounding organs. The effectiveness of this method was discussed in terms of sensitivity and accuracy. The segmented region GLCM, statistical, and morphological features were used to classify the tumor as benign or malignant.

Xiaoming et al. [28] proposed segmentation method to segment the liver out of from abdominal CT using a CNN-based framework. This method had effectively resulted in reliable segmentation. The classification process was reliable even with low processing time.

As a consequence of Alahmer's [29] work on the approaches for recognizing liver tumors and separating healthy from tumor liver tissue, CAD systems that use artificial intelligence and graphics rendering have received a lot of interest. The radiologist can next utilize a material image retrieval (CBIR) approach to help characterize liver lesions after segmenting the lesions. Then, a unique feature vector based on high-level features is generated and fed to SVM for lesion segmentation.

Amita et al. [30] introduced effective delineation of lesions on liver CT images using watersheds Gaussian guided convolutional (WGDL) technique. After tumor segmentation, various textural characteristics were extracted from the segmented lesions. An automated deep learning model (DNN) was used to further calssify detected tumors into either Hepatocellular, hemangioma carcinoma (HCC) or metastatic carcinoma.

As a preliminary step in finding liver abnormalities, Nalin et al. [31] proposed a computer-aided deep learning methods for segmenting the liver and lesions from an abdominal CT scans. The segmentation was achieved by GA-ANN, while the classification was achieved by LTEM method.

Almotairi et al. [32] proposed a deep learning algorithm for liver segmentation using CT liver scans. Their method was built on the top of the method presented in their paper [36]. The classification level was replaced with a binary pixel-by-pixel layer to simplify the binary separation of medical scans. Their method was efficient in terms of training time, memory requirements and accuracy.

According to Ayesha and Ghous [33], image-processing techniques had been shown efficient for analysis of medical images. Their algorithm design gave radiologist an effective tool for identifying malignant from benign cases. CAD technology made automatic tumor segmentation and hepatic tumor localization possible.

A computer-aided diagnosis method for detecting hepatitis and carcinoma (HCC) was proposed by Akash et al. [34]. Their method might be used as a screening tool in medical image analysis. Early detection and treatment of cirrhosis can prevent it from developing into HCC. A segmentation method based on Convolutional networks (CNNs) were proposed Kang et al. [35] to automate liver segmentation across various imaging modalities. They used double U-Net Network and broke training process into two epochs.

In order to employ the optimal classifier for CT liver scans, Samreen et al. [36] provided an experiment using machine learning classifiers such as MLP, SVM, RF, and J48. The results of the used classifiers were acceptable, but the MLP classifier performed better than all of them.

A fully automated method for segmenting liver tumors using CT scans was proposed by Jose et al. [37]. For this, two models were used; the original model was CNN with U-nit as the second model. The process of segmentation refinement decreased false positives by filling segmentation gaps.

Shaikh et al. [38] developed a CAD file liver cancer detection normalized hybrid features. The preprocessing step included median filtering, followed by adaptive threshold-based binary segmentation and ROI extraction utilizing morphological functions. From the ROI, textured features were extracted, combined, and normalized. The simulation results indicated that the proposed approach increased detection accuracy while minimizing processing power needed by classification stage.

Weiwei et al. [39] proposed an effective semi-automatic CAD system based on graph cuts and enhanced fuzzy means (FCM). The tumor volume of interest (VOI) was retrieved to decrease computing costs using a confidence-connected region growth technique. A kernelized FCM with spatial information was introduced into the graph cut segmentation process to enhance segmentation accuracy. The experimental results demonstrated efficacy of 3D liver tumors segmentation along with low processing time.

A brand-new two-stage liver detection and segmentation DSL model was proposed by Tang et al. [40]. Improved Faster Areas with CNN characteristics (Faster R-CNN) are used in the first stage to locate the liver's approximate location. To get the contour of the liver, the acquired pictures are processed and entered into Deep Lab. According to experimental data, the suggested method beats cutting-edge approaches in terms of average surface distance relative volume difference, volume overlap error, and overall score.

An efficient approach for segmenting liver tumors using graph cuts and adaptive region growth was developed by Zhen et al. [41]. First, adaptive region growth is used to extract the liver tumors as tumors were the regions of interest (ROIs) A manual seed is supplied for each tumor region. The ROIs are then improved via nonlinear mapping with Gaussian fitting based on the intensity distributions of the segmented tumor.

A 2.5D neural network was introduced by Girindra et al. [42]. Since the 2.5D model had a deeper and larger network design while still accommodating 3D information, its usage had shown encouraging results. The performance of the network and network parameter setup were related.

By representing multiscale global and local features at a finer level, Devidas and Sanjay [43] developed a multiscale technique to enhance the quality of CNN segmentation. They adjusted the channel-wise responses of the aggregate multiscale features to improve the network capabilities for handling high-level features. The experimental outcomes showed that the suggested model using 3Dircadb dataset was effective. The multiscale technique was able to lower the computational complexity while enhancing network segmentation performance.

According to Lang et al. [44], the massive development of AI, precision diagnostic and treatment systems along with public awareness of liver cancer have all led to improvements in the disease detection and management. AI-based computer technology has been applied to diagnosis, detection, treatment, and rehabilitation stages of liver cancer. Patients now have more tailored treatment options and possibilities for recovery thanks to machine learning and deep learning.

Amitha and Jayasree [45] proposed a liver and tumor segmentation technique. It was advantageous by having high noise robustness due to the usage of MRF integrated level set mechanism. Their method was able to eliminate shape ambiguity, precisely segment tumors and make organ measurements and visualization more accessible to radiologists and surgeons. After successful segmentation, feature extraction was carried out to classify tumors using SVM classifier.

Mala et al. [46] proposed a method to differentiate between hepatocellular carcinoma and cholangial carcinoma using abdominal CT images based on the probabilistic neural network (PNN). PNN was trained to identify the tumors using the textural features. Radiologists assisted in the evaluation of the results. This method was clinically applicable with a fair amount of precision.

AI based method was presented by Daniel et al. [47]. A CAD system was proposed for automatic categorization of localized liver lesions in CT scans. Hepatocellular carcinoma and liver cysts are classified using texture features analysis. The best descriptive feature which represents the typical deviation of horizontal curvature derived from the initial pixel grey levels, was used to fulfill the goal. This encouraging outcome paves the way for future expansion of this method to distinguish other forms of liver disorders using CT scans.

Mubasher et al. [48] proposed a method to identify and detect the tumor early using CT scans. The proposed study focuses on three machine learning (ML) techniques for multiclass liver tumor classification: logistic model tree (LMT), random tree (RT), and random forest (RF) with multiple automated regions of interest (ROI). Hemangioma, cyst, hepatocellular carcinoma, and metastasis are the four tumor classifications represented in the dataset. The CT scans were changed to grayscale, and histogram equalization was used to enhance their contrast. An automated system for hepatic tumor diagnosis using abdominal CT scans was presented by George et al. [49]. The tumors regions segmentation along with the process of the liver segmentation was automated using thresholding method. Their method was able to detect all malignancies while reducing the liver segmentation errors.

Despite the major advancements and improvements in the liver-abdomen CT image analysis, there were some significant concerns. First, it is noise sensitivity. Segmentation methods with low noise robustness and high noise sensitivity might result in over or under-segmentation and the inability to outline the liver and tumors accurately. Second, the liver and tumors exist in various shapes through CT slices. Hence, it might impose high computation complexity in the training stage to learn different liver shapes through CT slices.

The proposed method was able to overcome those challenges as follows. First, the proposed method was able to overcome segmentation noise sensitivity by applying HU scaling to focus on the liver intensities only and reduce the effect of impulse noise using edge-enhancing diffusion (EED) filter, which had the ability to denoising CT scans without affecting the scan structure, improving the perceptibility of the liver CT scan using color mapping to enhance the contrast between the liver intensities and the background intensities. Second, the 3D CNN DeepMedic network model with two pathways supplied with two versions of the liver- intensities focused preprocessed CT scans was able to reduce training computation complexity.

## 3. METHODOLOGY

The main concern of this work is to develop an automatic liver segmentation framework to detect and segment the liver out of CT abdomen scans. This work's main idea depends on using a 3D CNN network with multiple paths. Unlike the original 3D CNN with only a single path of input, the proposed 3D CNN architecture has two input pathways. This will be beneficial in improving the quality of the segmentation. The paper's main objective is to segment the liver out of other abdomen organs accurately. The accurate liver segmentation will be beneficial in reducing error rates produced by tumors mislabeled as liver tissue and vise-versa. This is mainly because liver segmentation is used as a pre-stage of tumor segmentation to reduce false segmentations resulting from segmenting some abdomen regions as tumors. Instead of passing the whole abdomen CT image to the tumor segmentation algorithm, only the liver region is passed to the algorithm. Hence, the higher the quality of liver segmentation, the lower the false tumor segmentation rates. Figure (1) describes the main architecture of the proposed method.

The proposed framework has two main stages: CT scans preprocessing and the actual liver segmentation (i.e., ROI extraction). Figure (2) illustrates the main architecture of the used 3D CNN network for liver detection and segmentation. It could distinguish the liver region from other abdomen organs with reasonable accuracy. The CT scans are first preprocessed then two versions of preprocessed CT scans are passed to the 3D CNN network for the segmentation task. The proposed system stages are discussed in the following subsections in detail.

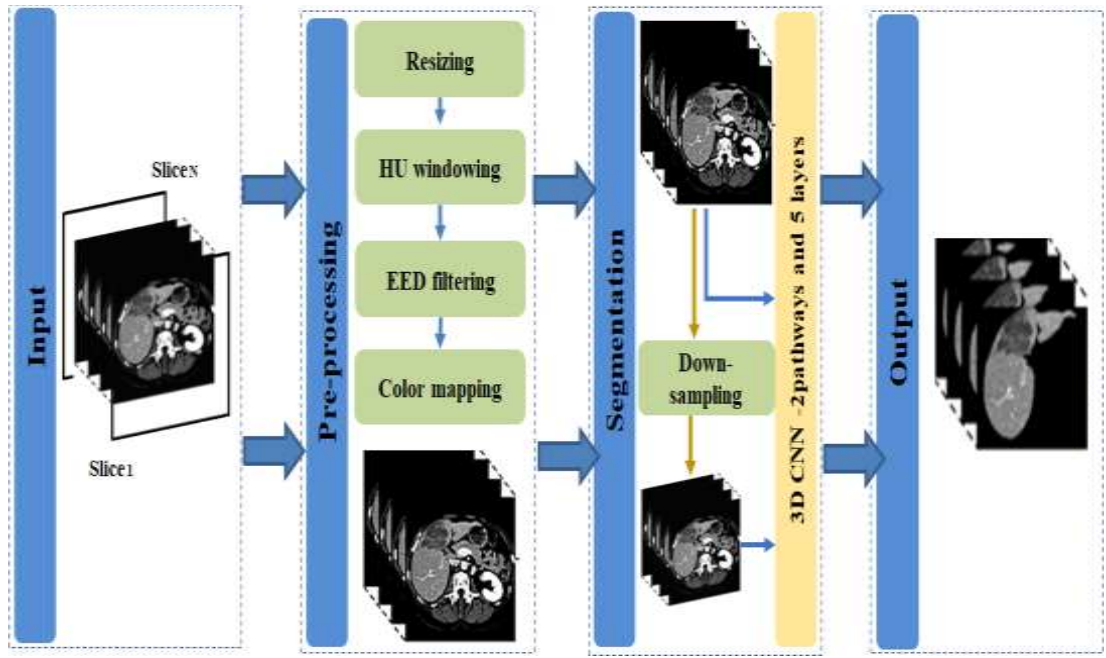

Figure (1). The proposed system architecture.

## 3.1. DATASETS

3D CT volumes are created using a CT scanner with rotating X-rays around the depicted body region. Each CT volume is associated with a single subject and is composed of several 2D images or slices of the depicted region. In this work, CT scans were acquired from three benchmark datasets, which are MICCAI-Sliver07 [50], LiTS17 [51], and 3Dircadb [52]. MICCAI-Sliver07 and LiTS17 CT scans are provided by ISBI challenges and in RAW format. 3Dircadb CT scans are provided by IRCAD institute in DICOM format. The total number of volumes per dataset is 30, 200, and 20 for MICCAI-Sliver07, LiTS17, and 3Dircadb, respectively. The number of slices (i.e., images) per CT volume varies through datasets. However, it can be limited to a certain range for each dataset as follows; MICCAI-Sliver07 slices vary in the range of 64 to 502 slices per volume, LiTS17 slices vary in the range of 42 to 1024 slices per volume, and 3Dircadb slices vary in the range of 74 to 260 slices per volume. The in-depth resolution of the used datasets slices is 512x512.

## 3.2. PRE-PROCESSING

The main objective of this stage is to enhance the perceptibility and the quality of the acquired CT scans while reducing the complexity of the segmentation task as possible. CT scans quality enhancement is achieved by reducing impulse noise and enhancing contrast. The quality of CT volumes in terms of slices images quality plays a critical role in enhancing the success rate of the segmentation process. To reduce the segmentation task complexity to match the limited capabilities of the CPU, CT slices were resized to the in-depth resolution of 265x265 instead of 512x512 resolution.

### 3.2.1. Clipping out ROI

After resizing the CT slices to the in-depth resolution of 265x265, the quality enhancement starts by enhancing the contrast between the liver (i.e., ROI) and the rest of the abdomen organs. This can be achieved by excluding abdomen organ intensities rather than the liver and tumor intensities using Hounsfield units (HU) windowing [53]. The liver intensities on the HU scale are in the range of [–200HU, 250HU]. Therefore, we used the HU window with values of [–100HU, 200HU] to cut out the liver intensities. This resulted in higher contrast CT abdomen images. This stage reduces the complexity of the segmentation task by focusing only on the ROI since the main concern of this paper is to automate the liver segmentation efficiently as a pre-stage for tumor segmentation while maintaining reasonable accuracy.

### 3.2.2. CT scans intensities normalization

Since the used CT scans are acquired from different datasets, the possibility of having various greyscales and imaging noises due to different scanning environments is relatively high. The existence of high greyscale variance may increase the training time. Hence, it increases the complexity of the segmentation task. To overcome such a problem, all CT scans were normalized to be in the range of [0,1] using Eq. (1).

$$I'(x,y,z) = \frac{I(x,y,z) - HU_{min}}{HU_{max} - HU_{min}} \qquad (1)$$

where $HU_{min}$ and $HU_{max}$ are the minimum and the maximum of the used HU range, respectively, which are –100 and 200. The value of 3D CT voxels before and after normalization is represented by $I(x,y,z)$ and $I'(x,y,z)$ respectively.

### 3.2.3. CT scans intensities color-mapping

For further enhancing the perceptibility of the preprocessed CT slices, normalized CT intensities are then color mapped to greyscale by multiplying them by 255. This is mainly because greyscale is in the range of [0, 255]. The more the contrast between ROI and other out-of-interest regions, the less the complexity of the segmentation task.

### 3.2.4. CT scans noise reduction:

The impulse noise imposed by the CT scanner can affect the quality of the segmentation process, and it may result in over or under-segmentation. CT impulse noise is reduced using edge-enhancing diffusion (EED) filter, which is considered an anisotropic diffusion filtering method [54]. The EED filter enhances image edges while reducing the noise effect. In other words, it preserves homogeneous structures while filtering the noise. The anisotropic diffusion process aims to equally diffuse concentration differences without destroying original input image structures or creating artifacts. EED was built on the Gaussian smoothing (GS) algorithm [55]. The general equation of anisotropic diffusion is given by

$$\frac{\delta I}{\delta t} = \nabla \cdot (D \cdot \nabla I) \qquad (2)$$

where $\nabla$ is the divergence operator, $\nabla I$ is the input image $I$ gradient, and $D$ is the diffusion variable by which the diffusion steers. The value of $\frac{\delta I}{\delta t}$ represents image $I$ intensities at time $t$. When a scalar-valued operator $g$ is used instead of D, the diffusion will be isotropic. The most commonly known isotropic filter is GS, in which $g$ is set to 1. Unlike GS, EED smooth image intensities in more than one direction at a time. The direction and magnitude of the smoothing process are determined by eigenvectors and eigenvalues, respectively. EED eigenvalues are defined by Eq. (3).

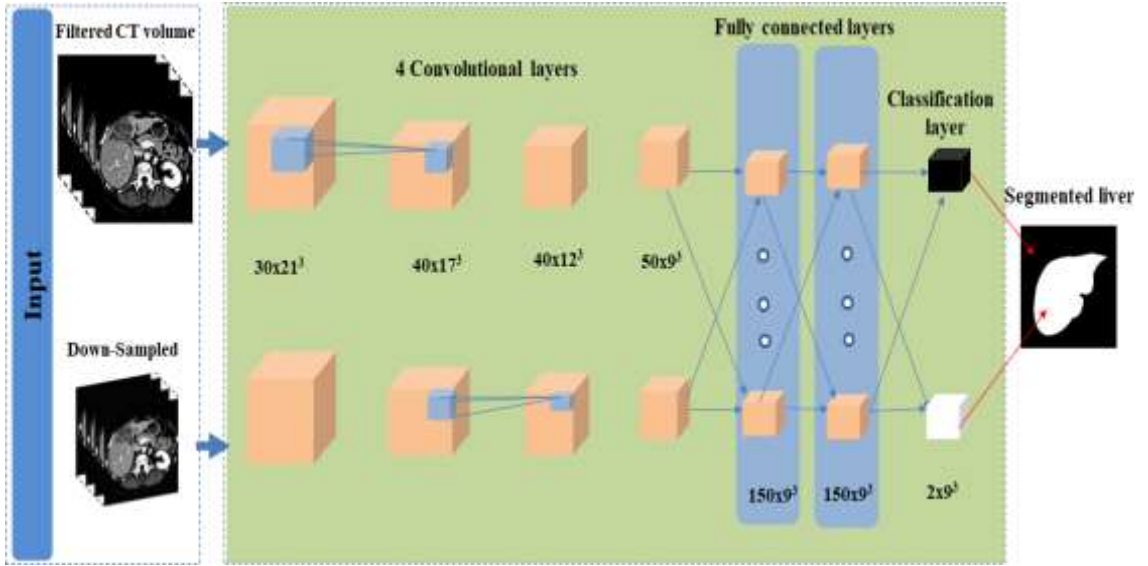

Figure (2). The proposed 3D CNN network settings.

$$\lambda_{e_1} = \begin{cases} 1, & (|\nabla I_s|^2 = 0) \\ \dfrac{-b}{\left(\dfrac{|\nabla I_s|^2}{\lambda_e}\right)^4} & (|\nabla I_s|^2 > 0) \end{cases} \quad (3)$$

$$\lambda_{e_2} = 1$$

where $\lambda_{e_1}$ and $\lambda_{e_2}$ are the eigenvalues of the EED, $b$ is a thresholding parameter and its value is set to 3.315, $|\nabla I_s|$ is the gradient magnitude of the image at a scale $s$ and $\lambda_e$ is a contrast parameter that indicate edges. The eigenvectors of EED are the same as those of the input image. This is mainly because the diffusion process preserves the local strictures of the input image.

### 3.3 LIVER SEGMENTATION

The main objective of this stage is to efficiently de-line the liver region boundaries (i.e., ROI) with reasonable accuracy. The output of this stage is the ROI along with its map $M_{ROI}$. The proposed 3D CNN network architecture has two pathways. Preprocessed CT volumes from the previous stage are supplied to the 3D CNN network pathways with two in-depth resolutions. The first pathway is supplied with preprocessed CT volumes with 265x265 resolution, while the second pathway is supplied with the same CT volumes of the first pathway down-sampled to 128x128 in-depth resolution. The 3D CNN network model used in this work is called DeepMedic [56].

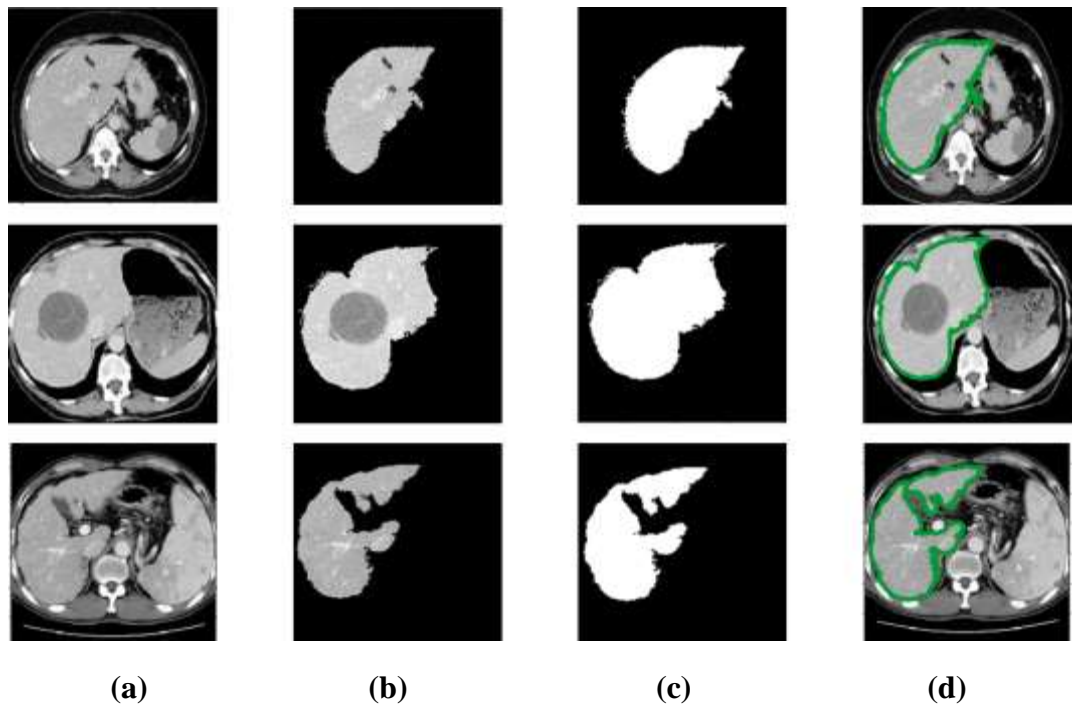

Figure (3). Liver segmentation results; (a) original input abdomen scan, (b) segmented liver, (c) segmented liver mask, and (d) the segmented liver outlined.

Unlike the original version of 3D CNN, which has a single pathway, the DeepMedic network architecture has two pathways. Multiple pathways are advantageous in training the network using a large set of local contextual information to segment the ROI area accurately. DeepMedic network has seven deep layers: four convolutional layers, two fully-connected layers, and a final classification layer. The final classification layer is a 3D fully connected conditional random field (CRF) layer. This layer enhances the quality of the overall network performance by reducing false rates resulting from mislabeling out-of-interest regions as ROI areas. CRF layer is advantageous by its ability to handle large neighborhoods in relatively short interference periods.

The segmentation problem using the DeepMedic network is highly dependable on the local and contextual information of each 3D CT voxel neighborhood to predict the final label of the voxel under processing. Convoluting CT scans of each pathway achieve the segmentation process with some filters at cascade convolutional layers. Those layers are groups of neurons that have the ability to extract and preserve certain patterns from the previous layer. Neurons activation is affected by the voxel receptive field as its increases in size through the subsequent layer.

The output of this stage is the liver segmented out of other abdomen organs along with $M_{ROI}$, which can be further used as input to tumor segmentation algorithms. Fig. 3 illustrates the experimental results of the proposed system.

## 4. RESULTS AND DISCUSSION

The proposed system was implemented on an HP machine using windows 10 under a processing environment of Intel Core I7-10[th] generation and 32 Gigabytes of RAM using MATLAB R2017a and Python. The applicability of the proposed system was evaluated using 250 CT volumes from three publicly available benchmark datasets, MICCAI-Sliver07, LiTS17, and 3Dircadb datasets. The CT volumes were divided into 70% for training, 10% for validation, and 20% for testing. The performance efficiency of the proposed system was evaluated using different performance metrics, which include accuracy (ACC), sensitivity (SEN), specificity (SPE), and the Dice similarity coefficient (DSC). ACC, SEN, SPE, and DSC can be calculated using the following equations:

$$ACC = \frac{TP + TN}{TP + TN + FP + FN} \tag{4}$$

$$SEN = \frac{TP}{TP + FN} \tag{5}$$

$$SPE = \frac{TN}{TN + FP} \tag{6}$$

$$DSC = \frac{2TP}{2TP + FP + FN} \tag{7}$$

where TP (true positive) is the count of pixels that were correctly labeled as ROI, TN (true negative) indicates the count of pixels rather than ROI that were correctly labeled as out-of-interest pixels, and FP (false positive) is the count of out of interest pixels that were mislabeled as ROI, and FN (false negative) is the count of all ROI pixels that were misclassified as out of interest pixels.

Table (1) illustrates the experimental results of the proposed system for each dataset, along with the average results for the system at all using all datasets. The results illustrated in Table (1) show that the proposed method achieved reasonable segmentation accuracy. Hence, the proposed method is an applicable method for liver segmentation. The high accuracy that the proposed method achieve comes as a consequence of not only 3D CNN network settings but also as a result of the preprocessing stage.

The preprocessing stage reduced the complexity of the segmentation problem. It eased the training process by first focusing on the intensities of the liver using HU windowing, enhancing the contrast between the liver and the rest of the abdomen organs, reducing the noise effect on the segmentation using EED, which can denoise the image while preserving its structures, reducing datasets intensities variance by normalizing all datasets CT scans to be in the range of [0, 1] and finally color mapping them to greyscale.

To highlight the effect of the preprocessing stage on the final segmentation quality, Table (2) illustrates the results of implementing the proposed 3D CNN network with un-preprocessed CT scans. Fig 4. illustrates the segmentation quality enhancement for the proposed system in terms of ACC, SEN, SPE, and DSC using preprocessed and unprocessed CT scans.

Table (1): Experimental results for the proposed 3D CNN model for liver segmentation with preprocessed CT scans.

| Dataset | ACC | SEN | SPE | DSC |
|---|---|---|---|---|
| **MICCAI-Sliver07** | 94.77% | 95.29% | 93.11% | 94.54% |
| **LiTS17** | 94.53% | 94.98% | 91.55% | 93.63% |
| **3Dircadb** | 93.78% | 93.44% | 90.93% | 91.26% |
| *Average* | *94.36%* | *94.57%* | *91.86%* | *93.14%* |

Table (2): Experimental results for the proposed 3D CNN model for liver segmentation with un-processed CT scans.

| Dataset | ACC | SEN | SPE | DSC |
|---|---|---|---|---|
| **MICCAI-Sliver07** | 81.92% | 82.07% | 79.89% | 81.32% |
| **LiTS17** | 81.53% | 82.88% | 81.45% | 81.53% |
| **3Dircadb** | 80.00% | 80.56% | 78.05% | 78.38% |
| *Average* | *81.15%* | *81.84%* | *79.80%* | *80.41%* |

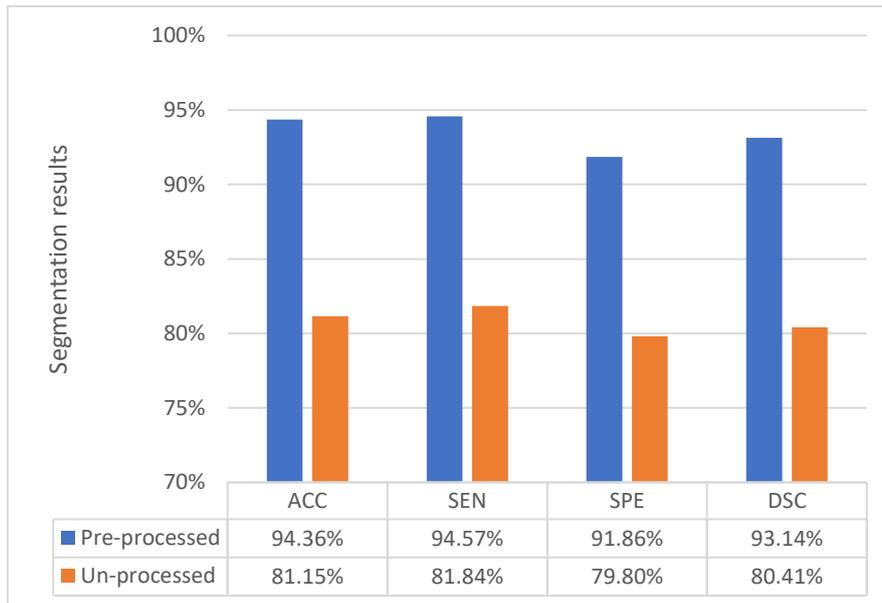

Figure (4): The effect of preprocessing CT scans on the quality of liver segmentation.

As illustrated in Fig. 4, the result segmentation quality using the proposed 3D CNN settings improved by almost 12% when the CT scans supplied to the network were preprocessed. The overall performance of the proposed system against other liver segmentation methods is presented in Table (3). From the results presented in Table (3), the proposed method achieved reasonable performance in terms of DSC. Hence, this proves the efficiency and applicability of the proposed method. The proposed method was able to segment the inner region of the liver with reasonable accuracy, but it suffers under segmentation on the boundaries of the liver.

Table (3): The proposed method against other segmentation methods in terms of DSC.

| Method | Liver Seg. DSC |
|---|---|
| Kaluva et al. [57] | 92.30% |
| Goryawala et al. [58] | 92.02% |
| Li et al. [59] | 92.23% |
| Huang et al. [60] | 92.50% |
| Xu et al. [61] | 93.00% |
| Moghbel et al. [62] | 91.10% |
| Y Al-Saeed et al. [63] | 92.40% |
| **The proposed Method** | **93.14%** |

## 5. CONCLUSION

This work presented an efficient technique for automatic liver segmentation to detect and segment the liver out of CT abdomen scans. It is based on the usage of 3D CNN with multiple paths. Unlike the ordinary 3D CNN with only a single path of input, the proposed 3D CNN architecture has two input pathways. It enables accurate liver segmentation out of other abdomen organs. Accordingly, the higher the quality of liver segmentation, the lower the false tumor segmentation rates. The proposed model was able to solve some of the open-research points related to the liver segmentation. First, high false rates resulted from mislabeling abdomen organs liver region due to high intensities overlap between the liver and the abdomen organs; which was solved by clipping out liver insanities using HU windowing. Then, problem of segmentation process being sensitive to noise which was solved by pre-processing, normalizing and color mapping the input CT scans before supplying them to the network. Finally, the problem raised by automatic liver segmentation high complexity in the training stage. This problem was solved by reducing segmentation problem complexity by focusing on the ROI rather than the whole abdomen CT scan using HU windowing. The proposed model achieved 94.36% accuracy and improved the experimental results in terms of the Dice similarity coefficient (DSC) to have a value of 93.14% better than the previously published work.

The future work will target improving the accuracy, sensitivity, and DSC for automatic segmenting liver out of abdomen scans and then the liver tumors out of the liver. Also, we plan to use this work's output to build a full computer-aided diagnosis system (CAD) for liver tumors diseases. The CAD system will use the output of this work as input to a tumor segmentation method. Then, the segmented tumors will be investigated using AI methods to label and diagnosis those tumors as malignant or benign tumors.

# REFERENCES

[1] J. C. Ozougwu, "Physiology of the liver," *International Journal of Research in Pharmacy and Biosciences,* vol. 4, pp. 13-24, 2017.

[2] H. Wang, X. Liang, G. Gravot, C. A. Thorling, D. H. Crawford, Z. P. Xu*, et al.*, "Visualizing liver anatomy, physiology and pharmacology using multiphoton microscopy," *Journal of biophotonics,* vol. 10, pp. 46-60, 2017.

[3] A. Snell, "Liver function tests and their interpretation," *Gastroenterology,* vol. 34, pp. 675-682, 1958.

[4] C. E. Cornelius, "The use of nonhuman primates in the study of bilirubin metabolism and bile secretion," *American Journal of Primatology,* vol. 2, pp. 343-354, 1982.

[5] E. J. Joo, Y. Chang, J. S. Yeom, and S. Ryu, "Hepatitis B virus infection and decreased risk of nonalcoholic fatty liver disease: a cohort study," *Hepatology,* vol. 65, pp. 828-835, 2017.

[6] Z. Younossi and L. Henry, "Systematic review: patient-reported outcomes in chronic hepatitis C-the impact of liver disease and new treatment regimens," *Alimentary pharmacology & therapeutics,* vol. 41, pp. 497-520, 2015.

[7] H. K. Seitz, R. Bataller, H. Cortez-Pinto, B. Gao, A. Gual, C. Lackner*, et al.*, "Alcoholic liver disease," *Nature reviews Disease primers,* vol. 4, pp. 1-22, 2018.

[8] S. Bellentani, "The epidemiology of nonalcoholic fatty liver disease," *liver international,* vol. 37, pp. 81-84, 2017.

[9] R. M. van Aerts, L. F. van de Laarschot, J. M. Banales, and J. P. Drenth, "Clinical management of polycystic liver disease," *Journal of hepatology,* vol. 68, pp. 827-837, 2018.

[10] F. Khan, "Automated segmentation of CT liver images: a review," *Journal of Communications Technology, Electronics and Computer Science,* vol. 19, pp. 5-9, 2018.

[11] D. Pescia, N. Paragios, and S. Chemouny, "Automatic detection of liver tumors," in *2008 5th IEEE International Symposium on Biomedical Imaging: From Nano to Macro*, 2008, pp. 672-675.

[12] J. C. Ahn, A. Connell, D. A. Simonetto, C. Hughes, and V. H. Shah, "Application of artificial intelligence for the diagnosis and treatment of liver diseases," *Hepatology,* vol. 73, pp. 2546-2563, 2021.

[13] D. Mathieu, V. Vilgrain, A.-E. Mahfouz, M. C. Anglade, M. P. Vullierme, and A. Denys, "Benign liver tumors," *Magnetic resonance imaging clinics of North America,* vol. 5, pp. 255-288, 1997.

[14] I. R. Wanless, "Benign Liver Tumours," *Sherlock's Diseases of the Liver and Biliary System,* pp. 671-680, 2011.

[15] D. Patel, Y. Shah, N. Thakkar, K. Shah, and M. Shah, "Implementation of artificial intelligence techniques for cancer detection," *Augmented Human Research,* vol. 5, pp. 1-10, 2020.

[16] N. Nishida and M. Kudo, "Artificial intelligence in medical imaging and its application in sonography for the management of liver tumor," *Frontiers in*

*Oncology,* vol. 10, p. 594580, 2020.

[17] W. Hu, H. Yang, H. Xu, and Y. Mao, "Radiomics based on artificial intelligence in liver diseases: where are we?," *Gastroenterology Report,* vol. 8, pp. 90-97, 2020.

[18] L.-Q. Zhou, J.-Y. Wang, S.-Y. Yu, G.-G. Wu, Q. Wei, Y.-B. Deng*, et al.*, "Artificial intelligence in medical imaging of the liver," *World journal of gastroenterology,* vol. 25, p. 672, 2019.

[19] Y.-L. Huang, J.-H. Chen, and W.-C. Shen, "Diagnosis of hepatic tumors with texture analysis in nonenhanced computed tomography images," *Academic radiology,* vol. 13, pp. 713-720, 2006.

[20] L. Massoptier and S. Casciaro, "A new fully automatic and robust algorithm for fast segmentation of liver tissue and tumors from CT scans," *European radiology,* vol. 18, pp. 1658-1665, 2008.

[21] A. M. Anter, A. T. Azar, A. E. Hassanien, N. El-Bendary, and M. A. ElSoud, "Automatic computer aided segmentation for liver and hepatic lesions using hybrid segmentations techniques," in *2013 Federated Conference on Computer Science and Information Systems*, 2013, pp. 193-198.

[22] P. Dankerl, A. Cavallaro, A. Tsymbal, M. J. Costa, M. Suehling, R. Janka*, et al.*, "A retrieval-based computer-aided diagnosis system for the characterization of liver lesions in CT scans," *Academic radiology,* vol. 20, pp. 1526-1534, 2013.

[23] S. Kumar, R. Moni, and J. Rajeesh, "An automatic computer-aided diagnosis system for liver tumours on computed tomography images," *Computers & Electrical Engineering,* vol. 39, pp. 1516-1526, 2013.

[24] G. Sethi and B. S. Saini, "Computer aided diagnosis system for abdomen diseases in computed tomography images," *Biocybernetics and Biomedical Engineering,* vol. 36, pp. 42-55, 2016.

[25] D. Edwin and S. Hariharan, "Liver and tumour segmentation from abdominal CT images using adaptive threshold method," *International Journal of Biomedical Engineering and Technology,* vol. 21, pp. 190-204, 2016.

[26] P. F. Christ, F. Ettlinger, F. Grün, M. E. A. Elshaera, J. Lipkova, S. Schlecht*, et al.*, "Automatic liver and tumor segmentation of CT and MRI volumes using cascaded fully convolutional neural networks," *arXiv preprint arXiv:1702.05970,* 2017.

[27] A. Das, P. Das, S. S. Panda, and S. Sabut, "Adaptive fuzzy clustering-based texture analysis for classifying liver cancer in abdominal CT images," *International Journal of Computational Biology and Drug Design,* vol. 11, pp. 192-208, 2018.

[28] X. Liu, S. Guo, B. Yang, S. Ma, H. Zhang, J. Li*, et al.*, "Automatic organ segmentation for CT scans based on super-pixel and convolutional neural networks," *Journal of digital imaging,* vol. 31, pp. 748-760, 2018.

[29] H. Alahmer, "Automated Characterisation and Classification of Liver Lesions From CT Scans," University of Lincoln, 2018.

[30] A. Das, U. R. Acharya, S. S. Panda, and S. Sabut, "Deep learning based liver cancer detection using watershed transform and Gaussian mixture model

techniques," *Cognitive Systems Research,* vol. 54, pp. 165-175, 2019.

[31] E. A. Babushkina, L. V. Belokopytova, A. M. Grachev, D. M. Meko, and E. A. Vaganov, "Variation of the hydrological regime of Bele-Shira closed basin in Southern Siberia and its reflection in the radial growth of Larix sibirica," *Regional Environmental Change,* vol. 17, pp. 1725-1737, 2017.

[32] M. Rovere, A. Mercorella, E. Frapiccini, V. Funari, F. Spagnoli, C. Pellegrini*, et al.*, "Geochemical and geophysical monitoring of hydrocarbon seepage in the Adriatic Sea," *Sensors,* vol. 20, p. 1504, 2020.

[33] A. A. Khan and G. B. Narejo, "Analysis of abdominal computed tomography images for automatic liver cancer diagnosis using image processing algorithm," *Current Medical Imaging,* vol. 15, pp. 972-982, 2019.

[34] A. Nayak, E. Baidya Kayal, M. Arya, J. Culli, S. Krishan, S. Agarwal*, et al.*, "Computer-aided diagnosis of cirrhosis and hepatocellular carcinoma using multi-phase abdomen CT," *International journal of computer assisted radiology and surgery,* vol. 14, pp. 1341-1352, 2019.

[35] K. Wang, A. Mamidipalli, T. Retson, N. Bahrami, K. Hasenstab, K. Blansit*, et al.*, "Automated CT and MRI liver segmentation and biometry using a generalized convolutional neural network," *Radiology. Artificial intelligence,* vol. 1, 2019.

[36] S. Naeem, A. Ali, S. Qadri, W. Khan Mashwani, N. Tairan, H. Shah*, et al.*, "Machine-learning based hybrid-feature analysis for liver cancer classification using fused (MR and CT) images," *Applied Sciences,* vol. 10, p. 3134, 2020.

[37] J. D. L. Araújo, L. B. da Cruz, J. L. Ferreira, O. P. da Silva Neto, A. C. Silva, A. C. de Paiva*, et al.*, "An automatic method for segmentation of liver lesions in computed tomography images using deep neural networks," *Expert Systems with Applications,* vol. 180, p. 115064, 2021.

[38] M. S. I. Turab and V. Kadam, "Liver Cancer Detection and Grading using Efficient Computer Vision Techniques," *Annals of the Romanian Society for Cell Biology,* pp. 1740-1755, 2021.

[39] W. Wu, S. Wu, Z. Zhou, R. Zhang, and Y. Zhang, "3D liver tumor segmentation in CT images using improved fuzzy C-means and graph cuts," *BioMed research international,* vol. 2017, 2017.

[40] W. Tang, D. Zou, S. Yang, J. Shi, J. Dan, and G. Song, "A two-stage approach for automatic liver segmentation with Faster R-CNN and DeepLab," *Neural Computing and Applications,* vol. 32, pp. 6769-6778, 2020.

[41] Z. Yang, Y.-q. Zhao, M. Liao, S.-h. Di, and Y.-z. Zeng, "Semi-automatic liver tumor segmentation with adaptive region growing and graph cuts," *Biomedical Signal Processing and Control,* vol. 68, p. 102670, 2021.

[42] G. Wardhana, H. Naghibi, B. Sirmacek, and M. Abayazid, "Toward reliable automatic liver and tumor segmentation using convolutional neural network based on 2.5 D models," *International journal of computer assisted radiology and surgery,* vol. 16, pp. 41-51, 2021.

[43] M. F. Fathi, A. Bakhshinejad, A. Baghaie, D. Saloner, R. H. Sacho, V. L.

Rayz*, et al.*, "Denoising and spatial resolution enhancement of 4D flow MRI using proper orthogonal decomposition and lasso regularization," *Computerized Medical Imaging and Graphics,* vol. 70, pp. 165-172, 2018.

[44] Q. Lang, C. Zhong, Z. Liang, Y. Zhang, B. Wu, F. Xu*, et al.*, "Six application scenarios of artificial intelligence in the precise diagnosis and treatment of liver cancer," *Artificial Intelligence Review,* vol. 54, pp. 5307-5346, 2021.

[45] A. Raj and M. Jayasree, "Automated liver tumor detection using Markov random field segmentation," *Procedia Technology,* vol. 24, pp. 1305-1310, 2016.

[46] K. Mala, V. Sadasivam, and S. Alagappan, "Neural network based texture analysis of liver tumor from computed tomography images," *International Journal of Medical and Health Sciences,* vol. 2, pp. 12-19, 2008.

[47] D. Smutek, A. Shimizu, L. Tesar, H. Kobatake, and S. Nawano, "Artificial Intelligence Methods Application in Liver Diseases Classification from CT Images," in *PRIS*, 2006, pp. 146-155.

[48] M. Hussain, N. Saher, and S. Qadri, "Computer Vision Approach for Liver Tumor Classification Using CT Dataset," *Applied Artificial Intelligence,* pp. 1-23, 2022.

[49] M. G. Linguraru, W. J. Richbourg, J. Liu, J. M. Watt, V. Pamulapati, S. Wang*, et al.*, "Tumor burden analysis on computed tomography by automated liver and tumor segmentation," *IEEE transactions on medical imaging,* vol. 31, pp. 1965-1976, 2012.

[50] B. Van Ginneken, T. Heimann, and M. Styner, "3D segmentation in the clinic: A grand challenge," in *MICCAI workshop on 3D segmentation in the clinic: a grand challenge*, 2007, pp. 7-15.

[51] P. Bilic, P. F. Christ, E. Vorontsov, G. Chlebus, H. Chen, Q. Dou*, et al.*, "The liver tumor segmentation benchmark (lits)," *arXiv preprint arXiv:1901.04056,* 2019.

[52] L. Soler, A. Hostettler, V. Agnus, A. Charnoz, J. Fasquel, J. Moreau*, et al.*, "3D image reconstruction for comparison of algorithm database: A patient specific anatomical and medical image database," *IRCAD, Strasbourg, France, Tech. Rep,* vol. 1, 2010.

[53] T. Razi, M. Niknami, and F. A. Ghazani, "Relationship between Hounsfield unit in CT scan and gray scale in CBCT," *Journal of dental research, dental clinics, dental prospects,* vol. 8, p. 107, 2014.

[54] A. M. Mendrik, E.-J. Vonken, A. Rutten, M. A. Viergever, and B. van Ginneken, "Noise reduction in computed tomography scans using 3-D anisotropic hybrid diffusion with continuous switch," *IEEE transactions on medical imaging,* vol. 28, pp. 1585-1594, 2009.

[55] A. E. Petropoulos, G. F. Vlachopoulos, S. G. Skiadopoulos, A. N. Karahaliou, and L. I. Costaridou, "Improving image quality in dual energy CT by edge-enhancing diffusion denoising," in *13th IEEE International Conference on BioInformatics and BioEngineering*, 2013, pp. 1-4.

[56] K. Kamnitsas, E. Ferrante, S. Parisot, C. Ledig, A. V. Nori, A. Criminisi*, et al.*, "DeepMedic for brain tumor segmentation," in *International workshop*

*on Brainlesion: Glioma, multiple sclerosis, stroke and traumatic brain injuries*, 2016, pp. 138-149.

[57] K. C. Kaluva, M. Khened, A. Kori, and G. Krishnamurthi, "2D-densely connected convolution neural networks for automatic liver and tumor segmentation," *arXiv preprint arXiv:1802.02182,* 2018.

[58] M. Goryawala, S. Gulec, R. Bhatt, A. J. McGoron, and M. Adjouadi, "A low-interaction automatic 3D liver segmentation method using computed tomography for selective internal radiation therapy," *BioMed research international,* vol. 2014, 2014.

[59] C. Li, A. Li, X. Wang, D. Feng, S. Eberl, and M. Fulham, "A new statistical and Dirichlet integral framework applied to liver segmentation from volumetric CT images," in *2014 13th International Conference on Control Automation Robotics & Vision (ICARCV)*, 2014, pp. 642-647.

[60] C. Huang, X. Li, and F. Jia, "Automatic liver segmentation using multiple prior knowledge models and free-form deformation," in *Proceedings of the VISCERAL challenge at ISBI, CEUR workshop proceedings*, 2014, pp. 22-24.

[61] Z. Xu, R. P. Burke, C. P. Lee, R. B. Baucom, B. K. Poulose, R. G. Abramson*, et al.*, "Efficient multi-atlas abdominal segmentation on clinically acquired CT with SIMPLE context learning," *Medical image analysis,* vol. 24, pp. 18-27, 2015.

[62] M. Moghbel, S. Mashohor, R. Mahmud, and M. I. B. Saripan, "Automatic liver segmentation on computed tomography using random walkers for treatment planning," *EXCLI journal,* vol. 15, p. 500, 2016.

[63] Y. Al-Saeed, W. A. Gab-Allah, H. Soliman, M. F. Abulkhair, W. M. Shalash, and M. Elmogy, "Efficient Computer Aided Diagnosis System for Hepatic Tumors Using Computed Tomography Scans," *Comput. Mater. Contin,* vol. 71, pp. 4871-4894, 2022.